\documentclass[twocolumn, prl, superscriptaddress,notitlepage]{revtex4-2}
\usepackage{graphicx}
\usepackage{physics}
\usepackage{epsfig}
\usepackage{color}
\usepackage{xspace}
\usepackage{array}
\usepackage{amsmath}
\usepackage{amsfonts}
\usepackage[dvipsnames]{xcolor}
\usepackage{colortbl}
\usepackage[breaklinks,colorlinks,linkcolor=blue,citecolor=blue,urlcolor=blue]{hyperref}

\allowdisplaybreaks

\makeatletter
\let\saved@frontmatter@maketitle\frontmatter@maketitle
\let\saved@frontmatter@title\frontmatter@title
\let\saved@frontmatter@author\frontmatter@author
\let\saved@frontmatter@date\frontmatter@date
\let\saved@frontmatter@thanks\frontmatter@thanks
\let\saved@frontmatter@affiliation\affiliation
\let\saved@frontmatter@email\email
\makeatother

\begin{document}
\title{Temperature-Controlled Resonance in a Heteronuclear Quantum Gas Mixture}

\author{Xiaoyi Yang}
\affiliation{MOE Key Laboratory for Nonequilibrium Synthesis and Modulation of Condensed Matter,
Shaanxi Province Key Laboratory of Quantum Information and Quantum Optoelectronic Devices, School of Physics,
Xi'an Jiaotong University, Xi'an 710049, China}

\author{Tianyu Xu}
\affiliation{MOE Key Laboratory for Nonequilibrium Synthesis and Modulation of Condensed Matter,
Shaanxi Province Key Laboratory of Quantum Information and Quantum Optoelectronic Devices, School of Physics,
Xi'an Jiaotong University, Xi'an 710049, China}

\author{Shengli Ma}
\affiliation{MOE Key Laboratory for Nonequilibrium Synthesis and Modulation of Condensed Matter,
Shaanxi Province Key Laboratory of Quantum Information and Quantum Optoelectronic Devices, School of Physics,
Xi'an Jiaotong University, Xi'an 710049, China}

\author{Zhigang Wu}
\email{wuzhigang@quantumsc.cn}
\affiliation{Quantum Science Center of Guangdong-Hong Kong-Macao Greater Bay Area (Guangdong), Shenzhen 508045, China}

\author{Ren Zhang}
\email{renzhang@xjtu.edu.cn}
\affiliation{MOE Key Laboratory for Nonequilibrium Synthesis and Modulation of Condensed Matter,
Shaanxi Province Key Laboratory of Quantum Information and Quantum Optoelectronic Devices, School of Physics,
Xi'an Jiaotong University, Xi'an 710049, China}
\affiliation{Hefei National Laboratory, Hefei 230088, China}

\begin{abstract}
Single-channel resonances are  fundamental processes in scattering of  atoms, yet their occurrence is largely incidental and lacks systematic control.  
In this Letter, we propose a mechanism to realize a continuously tunable single-channel resonance by controlling the temperature of the heteronuclear mixture. By extending the Casimir-like mediated interaction to finite temperature, we demonstrate that thermal smearing of the Fermi surface reshapes the effective potential between impurities, giving rise to a temperature-controlled resonance (TCR) over a wide parameter range. 
As a direct consequence, the resonance position shifts systematically with temperature variation, providing a clear experimental signature of this mechanism.
We further investigate the quench dynamics of a Bose gas immersed in a Fermi sea and demonstrate that the observed temperature-dependent loss features in recent experiments are consistent with the TCR mechanism.
Our results establish temperature as a simple and experimentally accessible control knob for single-channel resonances in ultracold quantum gases.
\end{abstract}
\maketitle

{\it Introduction---} 
In most textbook treatments of resonant scattering in quantum mechanics, one considers a variation of the interaction potential such that the bound state closest to threshold is tuned across it, resulting in a divergence of the scattering amplitude~\cite{LandauLifshitz}. While this single-channel threshold resonance is physically transparent and underpins the theoretical notion of universality in ultracold atomic physics, it is rarely realized in actual atomic collisions. In practice, the tuning mechanisms available in experiments, such as magnetic or optical fields, couple to internal degrees of freedom, thereby introducing an inherently multi-channel character. The most celebrated example of such a multi-channel resonance is Feshbach resonance~\cite{Inouye1998,Kohler2006,Chin2010}, which has become an indispensable
tool for manipulating interactions in ultracold atomic systems. Compared with Feshbach resonances, a key advantage of a single-channel resonance is that it introduces no additional length scale and therefore provides the cleanest route to exploring universal phenomena such as Efimov physics~\cite{barontini2009,Nakajima2011,Naidon2017} and the unitary Fermi gas~\cite{Bloch2008,Horikoshi2010,zwerger2011bcs}. In this sense, beyond its conceptual appeal, a tunable single-channel threshold resonance would constitute a valuable experimental tool that complements Feshbach-based controls.

The difficulty in realizing a single-channel resonance lies in the fact that the 
interatomic potential is not easily manipulated by external fields. Mediated interactions in ultracold heteronuclear mixtures, however, may provide a viable route to overcoming this limitation. In such systems, atoms of the same species experience not only the bare interatomic interaction but also an effective interaction mediated by the other species. These boson- or fermion-mediated interactions have been extensively studied
~\cite{Heiselberg2000,PhysRevA.61.053605,PhysRevA.66.023605,Klein2005,Fuchs2007,PhysRevA.78.013619,PhysRevA.85.063616,DeSpielman2014,Schecter2014,Kinnunen2015,Suchet2017,Dehkharghani2018,Camacho2018a,Camacho2018,Sowinski_2019,Mistakidis_2019,Reichert_2019,PhysRevLett.124.163401,Liao2020,Mistakidis2020,PhysRevA.103.063317,PhysRevA.103.L021301,PhysRevLett.126.123403,Chen2022,PhysRevLett.129.153401,Drescher2023,Santiago-Garcia_2023,Baroni2024,NRP2024,Shen2024,PhysRevA.110.033304,SciPostPhys.16.1.023,PhysRevA.109.023327,PhysRevA.110.030101,PhysRevResearch.7.023053} and are known to be tunable via the interspecies scattering length~\cite{DeSalvo2019,ArguelloLuengo2022,Patel2023,ChinRKKY2025}.
As a result, ultracold heteronuclear mixtures have emerged as a versatile platform for exploring a wide range of many-body phenomena, including polaron (impurity) physics~\cite{ospelkaus2006,olmos2011,spethmann2012,knap2012,Casteels2013,dasenbrook2013,massignan2014,cetina2015,scazza2017,Pascal2018,keiler2019,Perez-Rios2021,chen2021,Fujii2022,scazza2022,Petkovic2022,wang2022,becker2022,bighin2022,wang2023,wang2023a,wang2025}, unconventional superfluid pairing~\cite{Bijlsma2000,Viverit2002,Efremov2002,Matera2003,Wang2006,Wu2016,Midtgaard2016,Midtgaard2017,Kinnunen2018}, and Efimov physics~\cite{Tung2014,Pires2014,maier2015,sun2019}.
Recent experiments have revealed pronounced temperature-dependent effects in mediated interactions within ultracold mixtures~\cite{Chen2022,ChinRKKY2025}. Despite these observations, the principles underlying such temperature dependence remain largely unexplored.

\begin{figure}[t]
\centering
\includegraphics[width=0.45\textwidth]{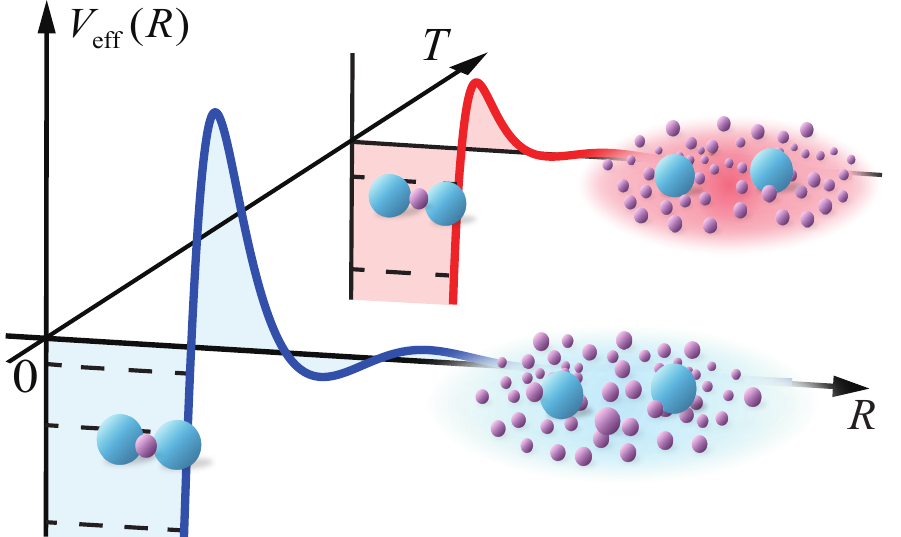}
\caption{Schematic of the Casimir-like mediated interaction $V_{\rm eff}(R)$ between two heavy impurities (blue) immersed in a Fermi sea of light fermions (purple). The interaction is mediated by the light fermions through contributions from both bound states and scattering states. 
The dashed lines indicate impurity–bound states supported by the mediated interaction.
The effective potential is continuously modified as the temperature $T$ is varied. }
\label{fig1}
\end{figure}

In this Letter, we propose a framework to achieve a continuously tunable 
single-channel resonance in a heteronuclear mixture by exploiting temperature as 
a control parameter. 
Our central finding is that the temperature can act as an effective tuning parameter for mediated interactions, driving the system across a single-channel resonance. We refer to this phenomenon as a temperature-controlled 
 resonance (TCR). We further validate this mechanism by studying 
the quench dynamics of a Bose-Fermi mixture, where our theoretical 
predictions show reasonable agreement with recent experimental 
measurements across a broad temperature range.


{\it Temperature-controlled single-channel resonance.-} We consider a two-component quantum mixture composed of heavy impurity particles (mass \(m_{\rm I}\)) and a light, single-component Fermi gas (mass \(m_{\rm F}\)), characterized by a mass ratio \(\eta = m_{\rm I}/m_{\rm F} > 1\). We assume that the impurities immersed within the Fermi sea are dilute and their bare interactions can be neglected. Our goal is to determine the finite-temperature effective potential \(V_{\text{eff}}(R,T)\) between a pair of such impurities,  mediated by the surrounding fermionic bath at an inter-impurity distance \(R\). Because of the large mass ratio (\(\eta \gg 1\)), we may treat this problem within the Born-Oppenheimer approximation. 
Specifically, the mediated interaction is determined by the grand potential of the Fermi gas, which includes contributions from both impurity--fermion bound states and the continuum of scattering states, as schematically depicted in Fig.~\ref{fig1}.

To calculate this interaction, we first consider the correction in the grand potential due to the presence of the two impurities, $\Delta \Omega(R,T)$, which is given by~\cite{Appendix}
\begin{align}
\label{eq:Tem-energy-correction}
\Delta \Omega(R,T) = &\Omega_{\rm b}(R,T) - k_{\rm B} T\notag\\ & \times\int_{0}^{\infty} \Delta\rho(E,R) \ln\left[1+e^{\beta(\mu-E)}\right] \, dE.
\end{align}
Here, $\beta = 1/(k_{\text{B}}T)$ is the inverse temperature, and $\mu$ is the chemical potential of the medium Fermi gas. The first term in Eq.~(\ref{eq:Tem-energy-correction}), $\Omega_{\rm b}(R,T)$, is the grand potential contribution from the bound states formed by a fermion and the two impurities. The second term represents the shift in grand potential coming from the continuum of scattering states, where $\Delta\rho(E,R)$ denotes the correction to the density of states (DoS) due to the impurity-fermion interaction. 
As in the zero-temperature case, both contributions are determined by solving the Schr\"odinger equation for the Fermi gas, treating the two impurities as static potential centers.

The impurity-fermion interaction is modeled using a contact potential, characterized by the $s$-wave scattering length $a_{\rm IF}$. The resulting Schr\"odinger equation reads $\left[ -(\hbar^{2}/2m_{\rm F})\nabla^{2} + V_{\rm IF}(\mathbf{r}) \right] \psi(\mathbf{r}) = E \psi(\mathbf{r})$, where the pseudopotential is defined as $V_{\rm IF}(\mathbf{r}) = (2\pi\hbar^2 a_{\rm IF}/m_{\rm F}) \left[ \delta(\mathbf{r} - \mathbf{R}/2) + \delta(\mathbf{r} + \mathbf{R}/2) \right]$ and $\psi({\bf r})$ represents the wave function of the fermion. For negative energies ($E < 0$), the system supports two bound states with even ($+$) and odd ($-$) parity. Their energies, $E_{\rm b,\pm}(R) = -\hbar^2 \kappa_{\pm}^2(R) / (2m_{\rm F})$, are determined by solving the transcendental equations, $\kappa_{\pm}(R) = a_{\rm IF}^{-1} \pm (1/R)e^{-\kappa_{\pm}R}$. The even-parity state corresponds to the ground state. In the single-component Fermi gas, each of these two states can be occupied by at most one fermion. Consequently, the bound-state contribution to the grand potential is $\Omega_{\rm b}(R,T) = - k_{\rm B} T \sum_{j \in \{+,- \}} \ln\left[ 1 + e^{\beta(\mu - E_{{\rm b},j}(R))} \right]$. The continuum scattering states are characterized by even- and odd-parity phase shifts, $\delta_{\pm}(k,R)$, given by $\tan\delta_{\pm}(k,R)=-(kR\pm\sin(kR))/(R a_{\rm IF}^{-1}\pm\cos(kR))$. The DoS correction for each channel is obtained using Friedel's sum rule, $\Delta\rho_{\pm}(E) = (1/\pi) d\delta_{\pm}/dE$. The total DoS correction is thus the sum over both channels:
\begin{equation}
\Delta\rho(E,R) = \frac{1}{\pi} \sum_{j \in \{+,-\}} \frac{d\delta_{j}(E,R)}{dE}.
\end{equation}

Having determined the total grand potential correction, we obtain the final effective potential by subtracting the asymptotic energy shift at infinite separation $R\to\infty$. This baseline, $\Delta \Omega(\infty,T)$, represents the independent energy contributions of two non-interacting impurities. Accordingly, the {\it temperature-dependent} effective potential is formally defined as
\begin{align}
\label{eq:Tem-potential}
V_{\rm eff}(R,T) = \Delta \Omega(R,T) - \Delta \Omega(\infty,T).
\end{align}
By construction, this potential is properly normalized to vanish at large distances ($V_{\rm eff} \to 0$ as $R \to \infty$) and naturally reduces to the established zero-temperature result in the $T \to 0$ limit~\cite{Nishida}.

To characterize the low-energy scattering properties dictated by $V_{\rm eff}(R,T)$, we evaluate the effective $s$-wave scattering length $a_{\rm eff}$. This quantity is determined from the scattering phase shift $\delta_{\rm eff}(k)$, which is governed by the variable-phase equation~\cite{calogero1967variable,Tilman2020}:
\begin{align}
k \partial_R \delta_{\rm eff}(k,R) = -m_{\rm I} V_{\rm eff}(R,T) \sin^2 \left[ kR + \delta_{\rm eff}(k,R) \right].
\end{align}
At short distances, the effective interaction diverges as $V_{\rm eff}(R,T) \sim \hbar^2/(m_{\rm F}R^2)$, presenting a strong singularity as $R\to0$. This unphysical divergence necessitates the introduction of a short-range cutoff radius $R_0$. 
Its specific value is not universal and will be determined by comparing the theoretical and experimental results~\cite{Appendix}. By integrating the variable-phase equation from this cutoff $R_0$ outwards to $R \to \infty$, we obtain the asymptotic phase shift $\delta_{\rm eff}(k) \equiv \delta_{\rm eff}(k, R\to\infty)$. Finally, the effective scattering length is extracted through the standard low-energy limit, $a_{\rm eff} = - \lim_{k\to0} \delta_{\rm eff}(k)/k$.

\begin{figure}[t]
\centering
\includegraphics[width=0.45\textwidth]{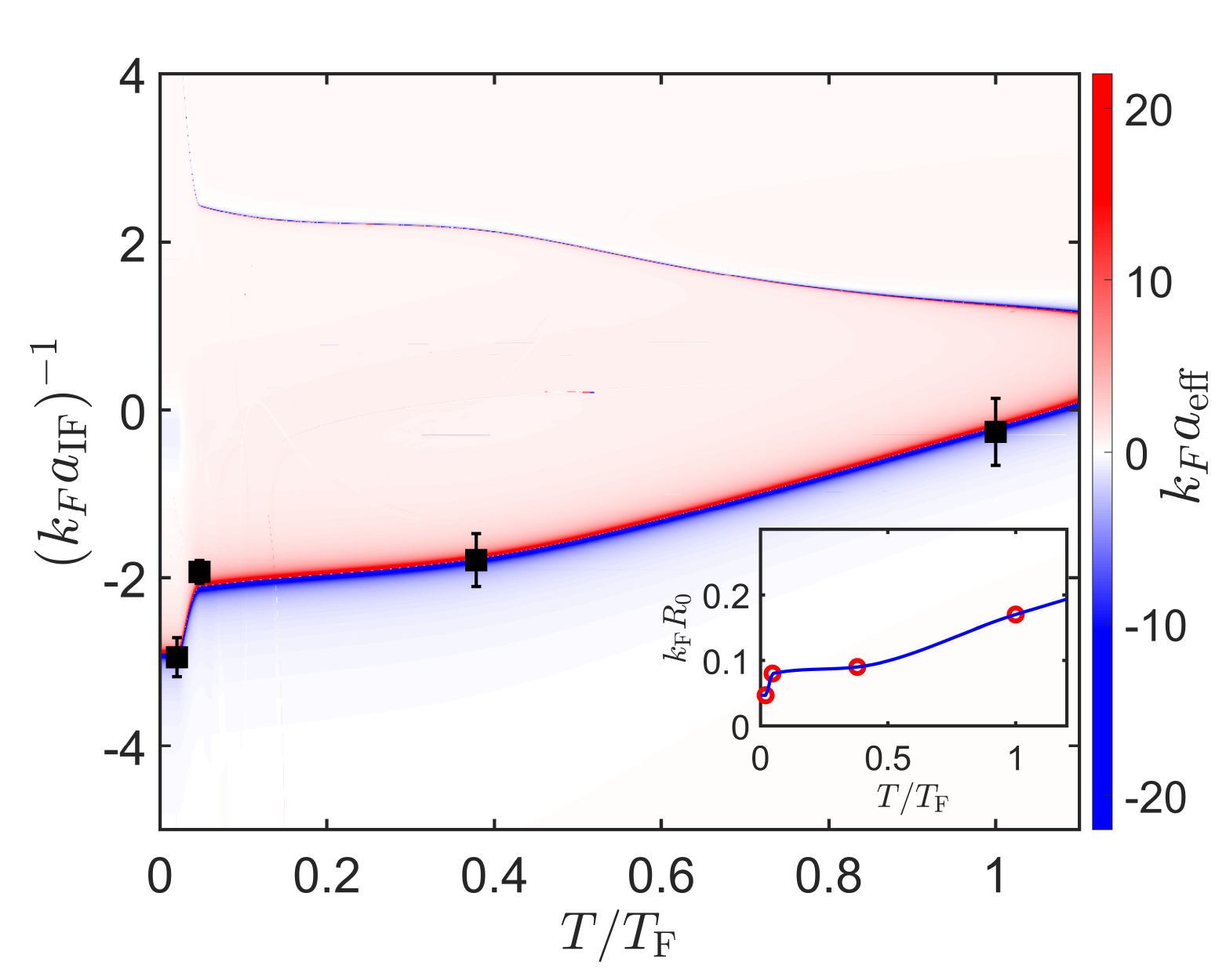}
\caption{
The effective scattering length $a_{\rm eff}$ as a function of the interspecies scattering length $a_{\rm IF}$ and temperature $T$, where the false color indicates values of $a_{\rm eff}$. The boundaries where $a_{\rm eff}$ transitions from large positive to large negative values correspond to the locations of resonances.
The markers indicate the experimentally measured loss centers of the Bose gas reported in Ref.~\cite{ChinRKKY2025}. 
The corresponding short-range cutoff values are $k_{\rm F}R_0=0.047$ at $T/T_{\rm F}\approx0.02$, $k_{\rm F}R_0=0.08$ at $T/T_{\rm F}\approx0.05$, $k_{\rm F}R_0=0.09$ at $T/T_{\rm F}\approx0.38$, and $k_{\rm F}R_0=0.17$ at $T/T_{\rm F}=1$, where $k_{\rm F}$ is the Fermi momentum.
The inset shows these cutoff values as a function of temperature, where the blue solid line is obtained by interpolating between them.
}
\label{fig2}
\end{figure}
To demonstrate that the variation of temperature can induce a resonance in the mediated interaction, we now consider a relevant experimental system, the Bose–Fermi mixture of a $^{133}$Cs–$^{6}$Li quantum gas. Shown in Fig.~\ref{fig2} is the effective scattering length $a_{\rm eff}$, determined using the above approach, plotted as a function of $a_{\rm IF}$ and $T$.  
We find that tuning the impurity--fermion scattering length $a_{\rm IF}$ at fixed temperature, or varying the temperature $T$ at fixed $a_{\rm IF}$, modifies the effective interaction $V_{\rm eff}(R,T)$. 
This leads to pronounced changes in the low-energy impurity--impurity scattering.
Indeed, $a_{\rm eff}$ exhibits clear divergences and sign changes under such tuning, signaling a resonance in the impurity scattering. These resonances constitute TCR, as they originate from temperature-dependent modifications of the effective interaction potential rather than from channel coupling.

Since resonances generally enhance atomic losses, TCR accounts for the experimentally observed loss pattern of impurity atoms as the temperature varies, with losses peaking at specific temperatures depending on the value of $a_{\rm IF}$~\cite{ChinRKKY2025}. The precise positions of the resonances--and thus of the loss peak--depend on the short-range cutoff $R_0$, which is not known {\it a priori} within our framework.  We therefore treat $R_0$ as a fitting parameter to reproduce the experimentally observed loss-peak locations (black markers in Fig.~\ref{fig2}). The short-range cutoff  $R_0$ so determined is shown in the inset of Fig.~\ref{fig2}, where we find that it is temperature-dependent, with higher temperatures corresponding to larger $R_0$.
\begin{figure}[t]
\centering
\includegraphics[width=0.45\textwidth]{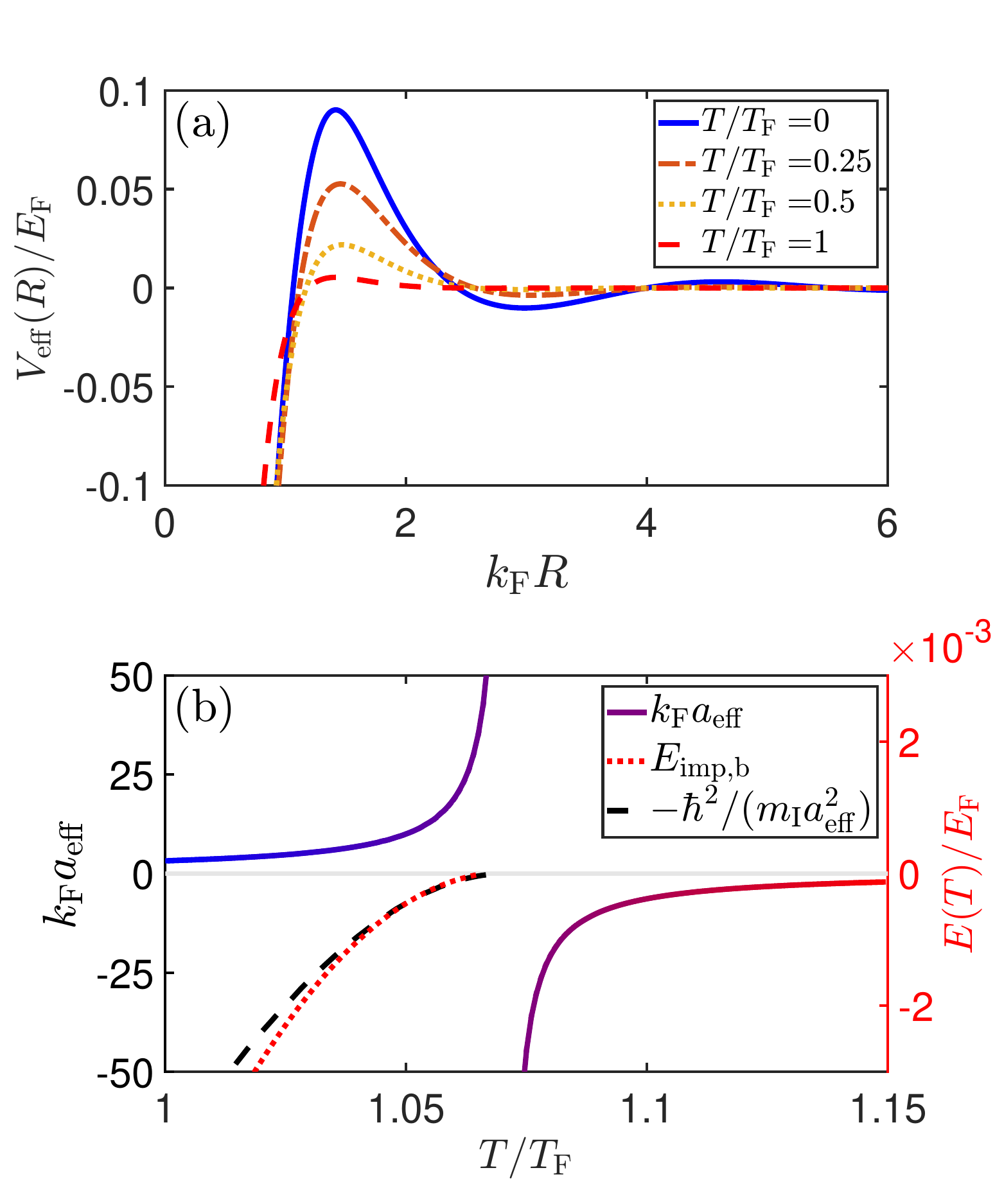}
\caption{(a) The mediated effective potential between two heavy impurities for the strongly interacting case $a_{\rm IF}\to\infty$.  (b) Effective scattering length $a_{\rm eff}$ as a function of temperature $T$ is shown on the left axis. The impurity bound-state energy $E_{\rm imp,b}$ (red dotted line) is shown on the right axis. The black dashed line indicates the low-energy prediction $-\hbar^2/(m_{\rm I}a_{\rm eff}^2)$. 
Energies are in units of Fermi energy $E_{\rm F}$ and temperatures in units of the Fermi temperature $T_{\rm F}$. } \label{fig3}
\end{figure}

\begin{figure*}[t]
\centering
\includegraphics[width=1\textwidth]{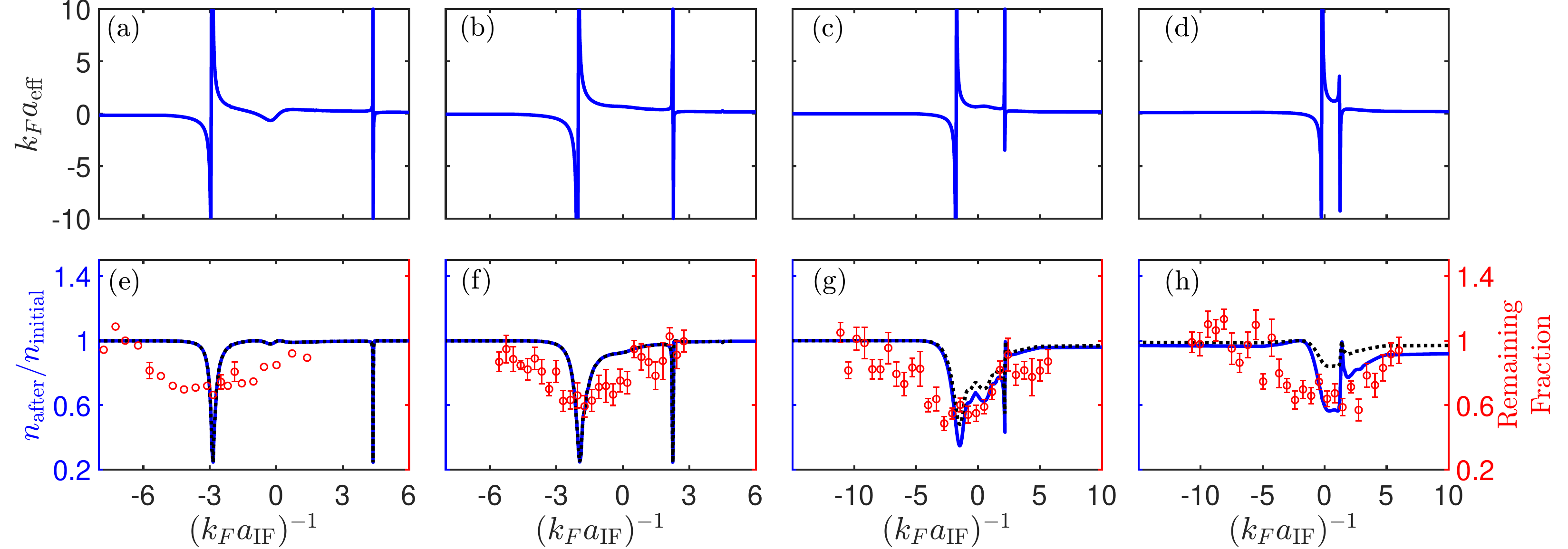}
\caption{
Comparison between the TCR prediction and quench dynamics at different temperatures.
(a)--(d) Effective scattering length $a_{\rm eff}$ as a function of the impurity--fermion scattering length $a_{\rm IF}$.
(e)--(h) Remaining low-momentum occupation $n_{\rm after}/n_{\rm initial}$ after the quench.
In panels (e)--(h), the left vertical axis denotes the calculated $n_{\rm after}/n_{\rm initial}$, while the right vertical axis denotes the experimentally measured loss signal from Ref.~\cite{ChinRKKY2025}.
The blue solid lines are obtained from the high-temperature virial expansion with local density approximation (LDA), and the black dotted lines show the corresponding results without LDA.
Red symbols represent the experimental data.
Panels (a) and (e) correspond to $T/T_{\rm F}\approx0.02$, (b) and (f) to $T/T_{\rm F}\approx0.05$, (c) and (g) to $T/T_{\rm F}\approx0.38$, and (d) and (h) to $T/T_{\rm F}=1$.
The cutoff $R_0$ is chosen using the same temperature-dependent values as in Fig.~\ref{fig2}.
}\label{fig4}
\end{figure*}

We now elucidate the mechanism of TCR by analyzing the strongly interacting regime defined by the condition $|a_{\rm IF}| \gg k_{\rm F}^{-1}, R_{0}$, where $k_{\rm F}$ is the Fermi momentum. In the zero-temperature limit, the effective potential admits an analytic form~\cite{Nishida}:
$
\frac{V_{\rm eff}(R,0)}{E_{\rm F}} \approx \frac{1}{\pi(k_{F}R)^{2}} \int_{0}^{(k_{F}R)^{2}} \tan^{-1} \left[ \frac{\sin(2\sqrt{x})}{x + \cos(2\sqrt{x})} \right] dx - \frac{W_{0}(1)^{2}}{(k_{\rm F}R)^2},
$ where $W_0(x)$ denotes the Lambert $W$ function. The first term corresponds to a long-range, Ruderman-Kittel-Kasuya-Yosida-type oscillatory interaction mediated by the sharp Fermi surface, while the second term originates from the bound states. 
At finite temperatures, where an analytic solution is unavailable, the effective potential is evaluated numerically according to Eq.~\eqref{eq:Tem-potential}. The resulting temperature-dependent potential profiles are displayed in Fig.~\ref{fig3}(a). As the temperature increases, the effective interaction is progressively suppressed. This suppression arises from the thermal broadening of the Fermi-Dirac distribution. Consequently, in the high-temperature limit, the effective potential is expected to disappear.

The resulting effective scattering length $a_{\rm eff}$ as a function of temperature is presented in Fig.~\ref{fig3}(b). As the temperature increases, the thermal modification of the effective potential drives the system across a resonance. This is manifested as a divergence in $a_{\rm eff}$ at a critical temperature $T_{\rm c} \approx 1.07 T_{\rm F}$, accompanied by a characteristic sign change. Physically, this divergence signals that a shallow impurity bound state supported by the effective interaction approaches zero energy. The binding energy $E_{\rm imp,b}$ of this bound state, obtained by solving the two-body Schr\"odinger equation with $V_{\rm eff}(R,T)$, is further shown with the red dotted line in Fig.~\ref{fig3}(b). 
Approaching the resonance, $E_{\rm imp,b}$ converges toward zero and shows excellent agreement with the universal low-energy prediction $ -\hbar^2/(m_{\rm I} a_{\rm eff}^2)$ (black dashed line). The simultaneous divergence of the scattering length and the vanishing of the binding energy serve as the definitive signatures of a scattering resonance. Consequently, the TCR mechanism provides a highly controllable framework for accessing and manipulating these temperature-dependent loss features in ultracold mixtures.

{\it Quench dynamics.-} 
In the experiment~\cite{ChinRKKY2025}, the loss of the impurity Bose gas was measured following a quench of the boson–fermion interaction strength in a Bose–Fermi mixture. At low temperatures, the observed loss features deviate from the positions associated with interspecies Feshbach and Efimov resonances. In addition, the loss center shifts systematically toward the unitary limit $a_{\rm IF}\to\infty$  as temperature increases. Our theoretical results account for the temperature-dependent shift in the loss features, as illustrated in Fig.~\ref{fig2}.

We further investigate the quench dynamics by employing the high-temperature virial expansion, which provides a framework for describing the change of the momentum distribution $n_{k}(t)$ of the impurity atoms following an interaction quench~\cite{virial,virial2}. In this approach, observables are expanded in powers of the fugacity $z$; keeping terms up to second order, one obtains~\cite{Appendix}
\begin{align}
n_{k}(t)=X_1z+\left(-Q_1X_1+X_2\right)z^2+{\cal O}\left(z^3\right). \label{eq:WExpressTo2}
\end{align}
Here, the expansion factor is $Q_m =\mathrm{Tr}_m [ e^{- \beta \hat{H}_0} ]$ and  
\begin{align}
X_m =\underset{\alpha,\gamma,\kappa}\sum e^{-\beta E_\alpha ^{(m)}}G_{\kappa \alpha} ^{(m)}(t)\langle \psi_{\kappa} ^{(m)}|\hat{n}_{k}|\psi_\gamma ^{(m)}\rangle G_{\gamma\alpha} ^{(m)}(t),
\end{align}
$m=1,2\cdots$ denotes the particle number, $E^{(m)}_\alpha$ and $\psi^{(m)}_\alpha$ respectively represent the energy and wave function of the $m$-particle non-interacting state, and $G^{(m)}(t)$ is the retarded Green's function of the $m$-particle interacting system. 
The virial expansion captures the key features of the quench dynamics: quenching from a non-interacting regime leads to a depletion of low-momentum occupations accompanied by an enhancement at higher momenta. 
Although this depletion is not identical to atom loss, both observables are expected to show enhanced responses when the quench approaches the resonance of the effective interaction. 
We therefore compare the position of maximal depletion with the experimentally observed loss centers, using their common temperature-dependent structure as a consistency check of the TCR mechanism.

The effective scattering length $a_{\rm eff}$ as a function of the interspecies scattering length $a_{\rm IF}$ at different temperatures is shown in Fig.~\ref{fig4}(a)--(d). The corresponding remaining boson fraction $n_{\rm after}/n_{\rm initial}$ after the quench is represented by black dotted lines in Fig.~\ref{fig4}(e)--(h). Here, $n_{\rm initial}$ and $n_{\rm after}$ denote the number of bosons in the lowest-momentum mode before the quench and at long times after the quench, respectively. To account for the trapping potential, we employ the local density approximation (LDA), shown by the blue solid lines. While LDA modifies the overall magnitude of the depletion, it does not shift the positions of the maximal depletion. 
In all cases, the largest depletion of the lowest-momentum occupation occurs when the system is quenched close to the strong interaction limit ($a_{\text{eff}} \to \infty$). This indicates that quenches to this regime inject the strongest excitation, resulting in the most pronounced redistribution of momentum. In contrast, quenches away from resonance generate weaker excitations and correspondingly smaller changes in the momentum distribution. The positions of maximal depletion lie close to the experimentally observed loss centers, indicating that these two distinct observables exhibit enhanced responses in the same interaction regime set by the TCR.

In addition, TCR predicts a narrow resonance feature on the repulsive side $a_{\rm IF}>0$ across all temperatures considered. 
We highlight Figs.~\ref{fig4}(f) and (g), where the experimental data show a weak but discernible enhancement in the loss signal near the predicted resonance positions.
Taken together, the proximity between the experimental loss features and the predicted resonance positions, along with their similar temperature dependence, supports TCR as a consistent framework.


{\it Conclusions.-}
We have identified a temperature-controlled  resonance (TCR) in heteronuclear mixtures. This mechanism arises from thermal modification of the Fermi sea, which reshapes the mediated effective interaction between impurities, allowing the resonant scattering to be tuned without changing the impurity-fermion scattering length.
In particular, the resonance position shifts systematically with temperature toward the strong interaction limit, consistent with experimental observations.
Using the high-temperature virial expansion, we analyze the quench-induced momentum redistribution and identify clear temperature-dependent features. The low-momentum depletions concur with the experimentally observed loss centers, both underpinned by the TCR. More generally, the TCR establishes temperature as a simple and broadly applicable tuning axis for many-body–induced interactions. This mechanism is not restricted to the specific Bose–Fermi mixture studied here, and is expected to apply generally to impurity problems in quantum gases, opening new opportunities for exploring tunable correlations and non-equilibrium dynamics across a wide range of systems.

\begin{acknowledgements}
We acknowledge fruitful discussions with Geyue Cai, Tao Chen, Bo Li, and Jikun Xie. This work is supported by NSFC (Grants Nos. 12574299, 12174300, 12474264), the Innovation Program for Quantum Science and Technology (Grant No. 2021ZD0302001), the National Key R\&D
Program of China (Grant No. 2022YFA1404103), and the Guangdong Provincial Quantum Science Strategic Initiative (Grant No.
GDZX2404007).
\end{acknowledgements}


%


\onecolumngrid
\newpage
\setcounter{equation}{0}
\setcounter{figure}{0}
\setcounter{table}{0}
\renewcommand{\theequation}{S\arabic{equation}}
\renewcommand{\thefigure}{S\arabic{figure}}
\renewcommand{\thetable}{S\arabic{table}}

\makeatletter
\frontmatter@init
\let\maketitle\saved@frontmatter@maketitle
\let\title\saved@frontmatter@title
\let\author\saved@frontmatter@author
\let\date\saved@frontmatter@date
\let\thanks\saved@frontmatter@thanks
\let\affiliation\saved@frontmatter@affiliation
\let\email\saved@frontmatter@email
\makeatother

\title{Supplementary material for ``Temperature-Controlled Resonance in a Heteronuclear Quantum Gas Mixture''}

\author{Xiaoyi Yang}
\affiliation{MOE Key Laboratory for Nonequilibrium Synthesis and Modulation of Condensed Matter,
Shaanxi Province Key Laboratory of Quantum Information and Quantum Optoelectronic Devices, School of Physics,
Xi'an Jiaotong University, Xi'an 710049, China}

\author{Tianyu Xu}
\affiliation{MOE Key Laboratory for Nonequilibrium Synthesis and Modulation of Condensed Matter,
Shaanxi Province Key Laboratory of Quantum Information and Quantum Optoelectronic Devices, School of Physics,
Xi'an Jiaotong University, Xi'an 710049, China}

\author{Shengli Ma}
\affiliation{MOE Key Laboratory for Nonequilibrium Synthesis and Modulation of Condensed Matter,
Shaanxi Province Key Laboratory of Quantum Information and Quantum Optoelectronic Devices, School of Physics,
Xi'an Jiaotong University, Xi'an 710049, China}

\author{Zhigang Wu}
\email{wuzhigang@quantumsc.cn}
\affiliation{Quantum Science Center of Guangdong-Hong Kong-Macao Greater Bay Area (Guangdong), Shenzhen 508045, China}

\author{Ren Zhang}
\email{renzhang@xjtu.edu.cn}
\affiliation{MOE Key Laboratory for Nonequilibrium Synthesis and Modulation of Condensed Matter,
Shaanxi Province Key Laboratory of Quantum Information and Quantum Optoelectronic Devices, School of Physics,
Xi'an Jiaotong University, Xi'an 710049, China}
\affiliation{Hefei National Laboratory, Hefei 230088, China}

\maketitle
\onecolumngrid
\setcounter{page}{1}
\setcounter{section}{0}
\setcounter{subsection}{0}

In this supplementary material, we present detailed derivations supporting the results discussed in the main text. It is organized into three sections: (1) Casimir-like interaction, (2) Induced phase shift, and (3) High-temperature virial expansion.

\section{1. Casimir-like interaction }

The effective interaction between impurities is mediated by the response of the surrounding Fermi gas. 
The presence of the impurities modifies the light-fermion spectrum, including bound states and the scattering continuum. 
At finite temperature, the spectral modification also changes the thermal occupation of fermionic states. 
Thus the mediated interaction is naturally obtained from the impurity-induced change in the grand potential of the Fermi gas, which includes both energy shifts and particle redistribution.
We write the grand potential shift induced by the two impurities as
\begin{align}
\Delta\Omega(R,T) = \Omega_{\rm b}(R,T) + \Delta\Omega_{\rm s}(R,T),
\end{align}
where $\Omega_{\rm b}(R,T)$ and $\Delta\Omega_{\rm s}(R,T)$ denote the contributions from bound states and scattering states, respectively.

We first evaluate the bound state contribution. 
For a large mass ratio $m_{\rm I}/m_{\rm F}\gg 1$, the heavy atoms can be treated as fixed scattering centers when solving the light-fermion problem. 
Consider a light fermion bound to two heavy impurities located at $\mathbf R/2$ and $-\mathbf R/2$. 
The bound-state wave function can be written as a superposition of two exponentially decaying waves centered at the two impurities,
\begin{align}
\psi_\pm(\mathbf r) = \frac{e^{-\kappa|\mathbf r - \mathbf R/2 |}}{|\mathbf r - \frac{\mathbf R}{2} |} \pm \frac{e^{-\kappa|\mathbf r + \mathbf R/2 |}}{|\mathbf r + \frac{\mathbf R}{2} |},
\end{align}
where the signs $\pm$ denote even and odd parity, respectively. The interaction between impurity and fermion is described by the contact potential, which is represented by the Bethe-Peierls boundary condition 
\begin{align}
\psi_\pm \left( \mathbf r \to \frac{\mathbf R}{2} \right) \propto \frac{1}{|\mathbf r - \frac{\mathbf R}{2}|} - \frac{1}{a_{\text{IF}}} + \mathcal O \left( |\mathbf r - \frac{\mathbf R}{2} | \right), 
\label{eq:BPcondition}
\end{align}
where $a_{\rm IF}$ is the $s$-wave scattering length between the fermion and the impurity. 
To determine the bound state energy $ - \hbar^2\kappa^2/2m_{\rm F}$, we impose this boundary condition on the wave function. 
Matching the short-distance behavior of $\psi_\pm(\mathbf r)$ gives
\begin{align}
\kappa_{\pm} = \frac{1}{a_{\text{IF}}} + \frac{1}{R} W_0(\pm e^{-R/a_{\text{IF}}}),
\end{align}
where $W_0$ is the Lambert function satisfying $z = W_0(z)  e^{W_0(z)}$. The bound state energy is $E_{{\rm b},j}(R) = - \hbar^2\kappa_\pm^2/2m_{\rm F}$, and its contribution to the grand potential is
\begin{align}
\Omega_{\rm b}(R,T) = - k_{\rm B}T \sum_{j \in \{ +,- \}} \ln \left[1+e^{-\beta(E_{{\rm b},j}(R)-\mu)}\right],
\end{align}
where the sum includes the parity channels that support physical bound states with $\kappa_j>0$.
For scattering states, the even- and odd-parity wave functions can be written as
\begin{align}
\psi_\pm(\mathbf r) =& \frac{\sin\left(k|\mathbf r - \mathbf R/2 | + \delta_\pm(E,R) \right)}{|\mathbf r - \frac{\mathbf R}{2} |} \pm \frac{\sin\left(k|\mathbf r + \mathbf R/2 | + \delta_\pm(E,R) \right)}{|\mathbf r + \frac{\mathbf R}{2} |},
\end{align}
where $\delta_\pm(E,R)$ is the phase shift.
To keep track of how the scattering potential modifies the density of states (DOS) of fermions, it is convenient to suppose that the system is confined in a large sphere with a radius $L\gg R$. For the even-parity states, the wave function near the boundary behaves as
\begin{align}
\psi_+(|\mathbf r|\to L) \sim \frac{2\sin\left(k L + \delta_+(E,R) \right) }{L}.
\end{align}
Thus the boundary condition $\psi_+(|\mathbf r|\to L)=0$ leads to the discretized momentum $k$,
\begin{align}
k L + \delta_+(E,R) = n \pi .
\end{align}
In the absence of impurities, this reduces to
\begin{align}
k L  = n \pi ,
\end{align}
where $n=1,2,3,\ldots$ labels the discretized scattering states. 
For a given momentum $k$, this quantization condition also gives the number of even-parity states with momentum below $k$. 
In the thermodynamic limit $L\to\infty$, the scattering spectrum becomes continuous. 
We denote by $N_+(E,R)$ the number of even-parity scattering states with energy below $E$ in the presence of the two impurities, and by $N_+^{(0)}(E)$ the corresponding number in the absence of impurities. 
Using $E=\hbar^2 k^2/(2m_F)$, the quantization conditions give
\begin{align}
N_+(E,R) = \frac{kL+\delta_+(E,R)}{\pi},
\qquad
N_+^{(0)}(E) = \frac{kL}{\pi}.
\end{align}
The change in the number of states is therefore
\begin{align}
\Delta N_+(E,R) \equiv N_+(E,R)-N_+^{(0)}(E)
= \frac{\delta_+(E,R)}{\pi}.
\end{align}
This result does not depend on the artificial boundary we introduced to discretize the spectrum.
We therefore use it to discuss the modification of the DOS.
The DOS is defined as
\begin{align}
\rho(E,R) = \frac{d N(E,R)}{d E},
\end{align}
which gives
\begin{align}
\Delta\rho_+(E,R) = \frac{d \Delta N_+(E,R)}{d E} = \frac{1}{\pi} \frac{d \delta_+(E,R)}{d E}.
\label{eq:DOSre}
\end{align}
At finite temperature, the resulting change of the continuum spectrum contributes to the grand potential with the thermal weight 
$-k_B T \ln[1+e^{-\beta(E-\mu)}]$. 
Therefore, the scattering contribution to the change of grand potential is
\begin{align}
\Delta \Omega_{\rm s,+}(R,T) = - k_{\rm B}T \int_0^\infty dE\, \Delta \rho_+(E,R) \ln \left[1+e^{-\beta(E-\mu)}\right] .
\end{align}
Substituting the change of DOS Eq.~(\ref{eq:DOSre}) and integrating by parts yields
\begin{align}
\Delta \Omega_{\rm s,+}(R,T) = k_{\rm B}T \frac{\delta_+(0,R)}{\pi} \ln \left[1+e^{\beta\mu}\right] - \int_0^\infty dE \frac{\delta_+(E,R)}{\pi} f(E).
\end{align}
In performing the integration by parts, we have used the fact that the phase shift vanishes in the high-energy limit, $\delta_+(E\to\infty)=0$. The lower boundary of the continuum gives the first term, which is determined by the zero-energy phase shift.  Only its $R$-dependent part contributes to the effective interaction, since any $R$-independent contribution is removed by subtracting the asymptotic value at $R\to\infty$. 
Including both even- and odd-parity channels, the grand potential shift of the scattering continuum is 
\begin{align}
\Delta \Omega_{\rm s} (R,T) = k_{\rm B}T \frac{\delta_+(0,R)+\delta_-(0,R)}{\pi} \ln \left[1+e^{\beta\mu}\right] - \int_0^\infty dk \frac{\hbar^2 k}{m_{\rm F}} \frac{1}{\pi}\bigl[\delta_+(k,R)+\delta_-(k,R)\bigr] f(k).
\end{align}
The phase shifts are determined by imposing the Bethe-Peierls boundary condition on the wave function, which gives
\begin{align}
\tan\delta_\pm(k,R) =  - \frac{kR \pm \sin(kR)}{\frac{R}{a_{\text{IF}}} \pm \cos(kR)}.
\end{align}

Combining this with the bound state contribution, the grand potential shift of the Fermi sea at finite temperature is
\begin{align}
\Delta \Omega(R,T) = \Omega_{\rm b}(R,T) + \Delta\Omega_{\rm s}(R,T).
\end{align}
When the two impurities are infinitely separated, their mediated interaction vanishes, and $\Delta\Omega(\infty,T)$ gives only the asymptotic single-impurity contribution. 
We therefore define the effective interaction between the two heavy atoms by subtracting this asymptotic value,
\begin{align}
V_{\rm eff}(R,T)=\Delta\Omega(R,T)-\Delta\Omega(\infty,T).
\label{eq:subtraction}
\end{align}
This leaves the fermion-mediated Casimir-like interaction that depends on the impurity separation $R$. 
This formulation reduces to the established zero-temperature result in the limit $T\to0$~\cite{SMNishida}. 

We now show this reduction explicitly by deriving the analytic form of the effective interaction in the zero-temperature and strong interaction limit. 
This limit also clarifies the two physical contributions to the effective interaction, namely the scattering continuum and the bound state. 
At zero temperature, for an occupied bound state with $E_b<\mu$, one has
\begin{align}
\lim_{T\to0} -k_{\rm B}T \ln \left[1+e^{-\beta(E_{\rm b}(R)-\mu)}\right] = E_{\rm b}(R)-\mu.
\end{align}
In the strong interaction limit $a_{\rm IF}\to\infty$, the bound state wave number is 
\begin{align}
\kappa_\pm = \frac{W_0(\pm 1)}{R}.
\end{align}
Only the even-parity channel supports a physical bound state in this limit, since $W_0(1)>0$, whereas $W_0(-1)$ is not real and therefore does not give a positive real decay constant.
The bound-state contribution to the grand potential is therefore
\begin{align}
\Omega_{\rm b}(R) = - \frac{\hbar^2 W_0(1)^2}{2m_{\rm F}R^2} - \mu .
\end{align}
The constant term $-\mu$ is independent of $R$ and is removed by the subtraction in Eq.~\eqref{eq:subtraction}.
At zero temperature, the Fermi-Dirac distribution becomes a step function, so that only scattering states below the Fermi surface contribute. 
The scattering part of the grand potential shift is then
\begin{align}
\Delta \Omega_{\rm s} (R,0) 
= - \int_0^{k_{\rm F}} dk \,
\frac{\hbar^2 k}{m_{\rm F}}
\frac{\delta_+(k,R)+\delta_-(k,R)}{\pi}.
\end{align}
Here the threshold term 
$k_{\rm B}T[\delta_+(0,R)+\delta_-(0,R)]\ln[1+e^{\beta\mu}]/\pi$ 
has been omitted because, in this limit, the zero-energy phase shift is independent of $R$ and is therefore removed in the definition of $V_{\rm eff}(R,T)$.

In the strong-interaction limit $a_{\rm IF}\to\infty$, the sum of the even- and odd-parity phase shifts is determined by
\begin{align}
\tan\left[\delta_+(k,R)+\delta_-(k,R)\right]
= - \frac{\sin(2kR)}{(kR)^2+\cos(2kR)} .
\end{align}
Substituting this expression into the scattering contribution gives
\begin{align}
\Delta \Omega_{\rm s} (R,0) 
= \frac{1}{\pi}
\int_0^{k_{\rm F}} dk \,
\frac{\hbar^2 k}{m_{\rm F}}
\tan^{-1}
\left[
\frac{\sin(2kR)}
{(kR)^2+\cos(2kR)}
\right].
\end{align}
Introducing the dimensionless variable $x=k^2R^2$, we have 
$dx=2kR^2dk$ and $k\,dk=dx/(2R^2)$. 
The upper limit $k=k_{\rm F}$ becomes $x=(k_{\rm F}R)^2$. 
Using $E_F=\hbar^2 k_F^2/(2m_{\rm F})$, the scattering contribution can be written as
\begin{align}
\frac{\Delta \Omega_{\rm s} (R,0)}{E_F}
= \frac{1}{\pi(k_{\rm F}R)^2}
\int_0^{(k_{\rm F}R)^2} dx \,
\tan^{-1}
\left[
\frac{\sin(2\sqrt{x})}
{x+\cos(2\sqrt{x})}
\right].
\end{align}

Combining this scattering contribution with the bound state contribution, and subtracting the asymptotic value at $R\to\infty$, we obtain the zero-temperature effective interaction in the strong interaction limit,
\begin{align}
\frac{V_{\mathrm{eff}}(R,0)}{E_F} 
= \frac{1}{\pi(k_{\rm F}R)^2}
\int_0^{(k_{\rm F}R)^2} dx \,
\tan^{-1}
\left[
\frac{\sin(2\sqrt{x})}
{x+\cos(2\sqrt{x})}
\right]
-\frac{W_0(1)^2}{(k_F R)^2}.
\end{align}
The first term comes from the redistribution of scattering states in the Fermi sea and gives an Ruderman-Kittel-Kasuya-Yosida-type oscillatory contribution.
The second term comes from the bound state and produces a short-range Efimov-type attraction proportional to $-1/R^2$.

\section{2.  Induced phase shift}

The previous section gives the fermion-mediated effective interaction $V_{\rm eff}(R,T)$ between the two heavy impurities. 
We now use this effective potential to extract the low-energy impurity--impurity scattering properties, in particular the $s$-wave phase shift and the effective scattering length $a_{\rm eff}$. 
For this purpose, we employ the variable phase method, following the standard scattering-theory formulation of Taylor~\cite{SMTaylor}. 
The key idea is to temporarily truncate the potential $V(R)$ at a radius $\rho$,
\begin{align}
V_\rho(R)=
\begin{cases}
V(R), & R<\rho,\\
0, & R>\rho.
\end{cases}
\end{align}
We denote the $s$-wave radial wave function for the actual potential $V(R)$ by $u(R)$ and the radial wave function for the truncated potential $V_\rho(R)$ by $u_{\rho}(R)$. The corresponding phase shift generated by $V_\rho(R)$ is denoted by $\delta(k,\rho)$, which approaches the actual phase shift $\delta_{\rm eff}(k)$ in the limit
\begin{align}
\delta(k,\rho)\xrightarrow[\rho\to\infty]{}\delta_{\rm eff}(k) .
\end{align}
Since the potential vanishes for $R>\rho$, the wave function in this region takes the free form
\begin{align}
u_\rho(R)=\alpha(\rho)\sin[kR+\delta(k,\rho)],
\end{align}
with amplitude $\alpha(\rho)$ and phase $\delta(k,\rho)$ to be determined. When $R\leq\rho$, $u(R)$ and $u_{\rho}(R)$ satisfy the same differential equation and the same short-range boundary condition. Thus, we have $u(R)\equiv u_\rho(R)$ at $ R = \rho$, and
\begin{align}
u(\rho)&=\alpha(\rho)\sin[k\rho+\delta(k,\rho)].
\label{Vu}
\end{align}
In the same way we obtain the derivative of $u(R)$ at $R = \rho$ as
\begin{align}
\partial_R u(\rho)&=k\alpha(\rho)\cos[k\rho+\delta(k,\rho)].
\label{Vu1}
\end{align}

\begin{figure}
\centering
\includegraphics[width=0.9\textwidth]{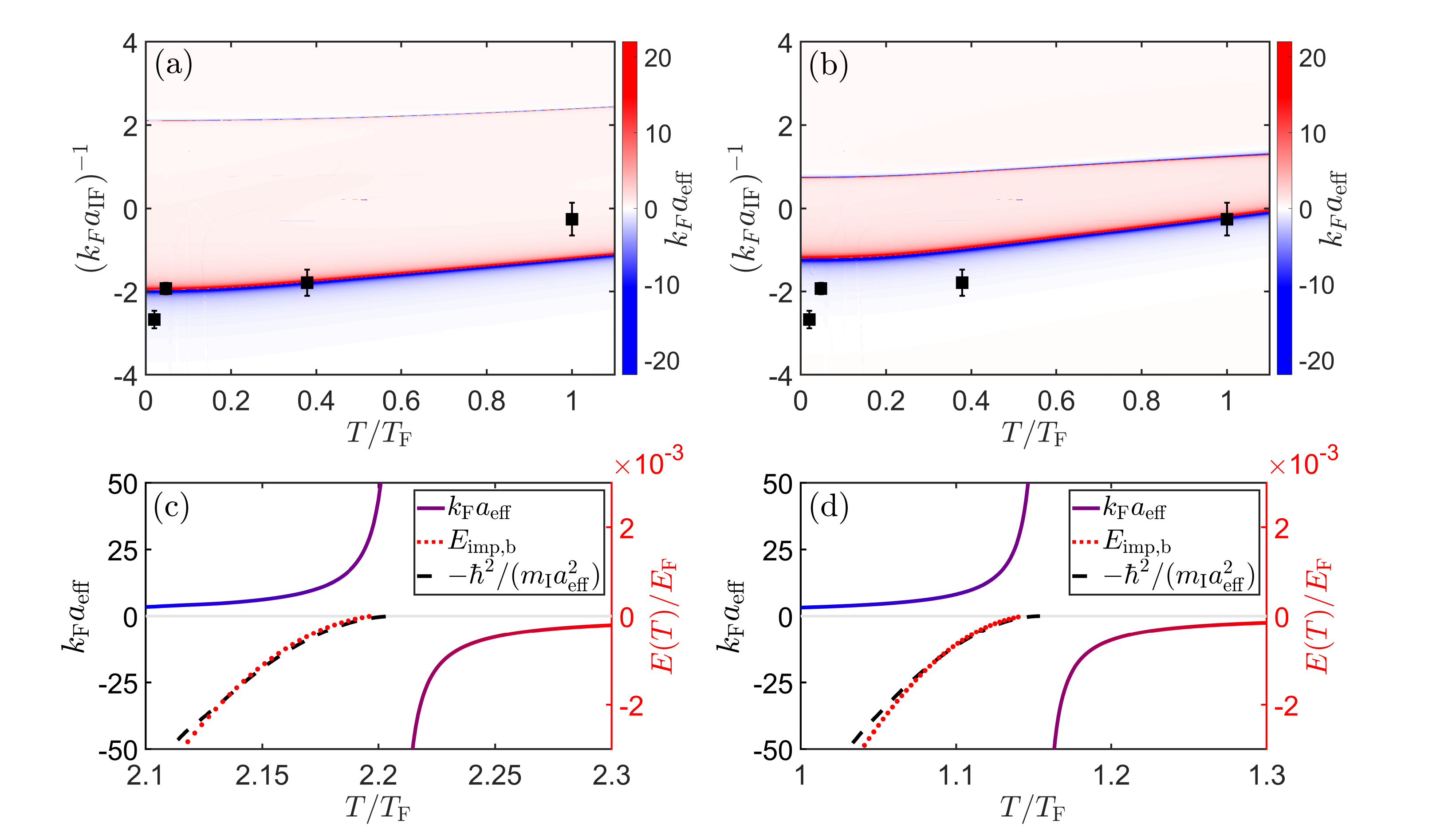}
\caption{(a) and (b) The effective scattering length $a_{\rm eff}$ as a function of the interspecies scattering length $a_{\rm IF}$ and temperature $T$. 
The markers indicate the experimentally measured loss centers of the Cs Bose gas reported in Ref.~\cite{SMChinRKKY2025}.
(c) and (d) The effective scattering length $a_{\rm eff}$ in the strong interaction limit $a_{\rm IF} \to \infty$ is shown on the left axis. The impurity bound-state energy $E_{\rm imp,b}$ (red dotted line) is shown on the right axis. The black dashed line indicates the low-energy prediction $-\hbar^2/(m_{\rm I}a_{\rm eff}^2)$. 
The short-range cutoff $k_{\rm F}R_0 = 0.09$ is fixed in (a) and (c). The cutoff in (b) and (d) is $k_{\rm F}R_0 = 0.17$.
}
\label{figS1}
\end{figure}


Since Eqs.~(\ref{Vu}) and (\ref{Vu1}) can be imposed at any truncation radius $\rho$, both the amplitude $\alpha(\rho)$ and the phase shift $\delta(k,\rho)$ change as the truncation radius is moved outward. 
We therefore treat $\rho$ as a continuous variable and replace it by $R$ in the following derivation. With this substitution, the second derivative of $u(R)$ from Eq.~(\ref{Vu1}) becomes 
\begin{align}
\partial_R^2 u(R) = k\left[\alpha'(R)\cos[kR+\delta(k,R)] -\alpha(R)[k+\delta'(k,R)]\sin[kR+\delta(k,R)]\right].
\label{Vu2}
\end{align}
Substituting Eqs.~(\ref{Vu}) and (\ref{Vu2}) into the radial Schr\"odinger equation, we obtain
\begin{align}
-k\frac{\alpha'(R)}{\alpha(R)}\cos[kR+\delta(k,R)]+k\delta'(k,R)\sin[kR+\delta(k,R)] = -m_{\rm I} V(R)\sin[kR+\delta(k,R)],
\label{RSE}
\end{align}
where $m_{\rm I}$ is the impurity mass in this work, and we set $\hbar=1$.
Requiring the derivative of $u(R)$ in Eq.~(\ref{Vu}) to be consistent with Eq.~(\ref{Vu1}) yields the constraint
\begin{align}
\alpha'(R)\sin[kR+\delta(k,R)] +\alpha(R)\delta'(k,R)\cos[kR+\delta(k,R)] =0.
\end{align}
Substituting this constraint into Eq. (\ref{RSE}), we obtain
\begin{align}
k\frac{d\delta(k,R)}{dR} =- m_{\rm I} V(R)\sin^2 [kR+\delta(k,R)].
\label{eq:variable-phase}
\end{align}
This is the variable phase equation for $s$-wave scattering. The initial condition is set by the short-range cutoff $R_0$. The short-range cutoff $R_0$ is introduced to account for the breakdown of the effective interaction at distances comparable to the range of the bare atomic interaction. It effectively encodes short-range physics not resolved by the long-range mediated interaction and should be regarded as a parameter characterizing unresolved short-distance processes. For a hard sphere of radius $R_0$, the wave function should vanish at the boundary $R=R_0$, i.e. 
\begin{align}
\alpha(R_0) \sin[k R_0 + \delta(k,R_0)] = 0. 
\end{align}
Thus, the boundary condition for the variable phase equation is $\delta(k,R_0)=-kR_0$. Integrating the variable phase equation to large $\rho\to\infty$ with the effective interaction $V_{\rm eff}(R)$ yields the actual effective phase shift $\delta_{\rm eff}(k)$. The effective scattering length is then obtained from the zero-energy limit, $a_{\rm eff}=-\lim_{k\to 0}\frac{\tan\delta_{\rm eff}(k)}{k}$.

Before using the temperature-dependent cutoff adopted in the main text, we first examine whether the temperature-controlled resonance persists for fixed short-range cutoffs. 
This comparison is intended to test the robustness of the resonance against the choice of $R_0$. 
Fig. \ref{figS1} shows the temperature dependence of $a_{\rm eff}$ for two fixed choices of the short-range cutoff $R_0$.
In both cases, the divergence of $a_{\rm eff}$ persists and shifts with temperature, showing that the temperature-controlled resonance is not tied to a particular cutoff. The comparison with the experimental markers, however, shows that a single fixed cutoff does not track the loss centers over the full temperature range.
Fig. \ref{figS1} (c) and (d) further examine the resonance structure and bound energy at strong interaction limit $a_{\rm IF}\to\infty$. The divergence of $a_{\rm eff}$ is accompanied by the impurity bound-state energy (red dotted line) approaching zero, consistent with the standard low-energy relation $E_{\rm imp,b}=-\hbar^2/(m_{\rm I}a_{\rm eff}^2)$ (black dashed line). 
These results indicate that the temperature-controlled resonance is driven by the temperature dependence of the effective interaction, rather than by a particular choice of the short-range cutoff $R_0$.

\section{3. High temperature virial expansion}


The previous sections establish the effective interaction mediated by the Fermi gas and the corresponding induced scattering length $a_{\rm eff}$. 
As a qualitative dynamical probe of the resonance structure, we consider a simplified interaction-quench problem.
The purpose of this calculation is not to compute the atom-loss rate directly, but to examine how strongly the low-momentum occupation is redistributed after a sudden change of the effective interaction.

We consider a thermal Bose gas initially in equilibrium with the free Hamiltonian $\hat H_0$ at temperature $T$. 
At $t=0$, the interaction is quenched from zero to a finite value characterized by the effective scattering length $a_{\rm eff}$. 
For $t>0$, the system evolves under the full Hamiltonian
\begin{align}
\hat H=\hat H_0+\hat V ,
\end{align}
where $\hat V$ denotes the post-quench interaction. 

The momentum occupation after the quench is evaluated with respect to the initial thermal ensemble, while the time evolution is governed by the post-quench Hamiltonian. It is given by
\begin{align}
n_{\bf k}(t)=
\frac{\mathrm{Tr}\left[
e^{- \beta (\hat{H}_0 -\mu \hat{N})}
e^{it\hat{H}} \hat{n}_{\bf k} e^{-it \hat{H}}
\right]}
{\mathrm{Tr}\left[e^{-\beta (\hat{H}_0 - \mu \hat{N})}\right]},
\end{align}
where $\beta=1/k_BT$ is the inverse temperature and $\hat N$ is the total particle-number operator of the Bose gas.

At high temperature, the fugacity $z=e^{\beta\mu}\ll 1$ is small. 
We then expand the partition function in powers of $z$. 
This fugacity expansion organizes the dynamics according to few-body sectors: the contribution from an $m$-particle sector appears at order $z^m$. 
In this regime, three-body and higher-body contributions are suppressed by higher powers of $z$ and are omitted in our calculation. 
Keeping terms up to second order in $z$, the momentum occupation takes the form
\begin{align}
n_{\bf k}(t)
=
X_1 z+\left(-Q_1X_1+X_2\right)z^2
+{\cal O}\left(z^3\right). 
\label{eq:WExpressTo2}
\end{align}
Here, the first-order term describes the single-particle contribution, while the second-order term contains the leading effect of interactions through the two-particle sector.

The coefficients $X_m$ and $Q_m$ are defined within the $m$-particle Hilbert space as
\begin{align}
X_m 
&=
\mathrm{Tr}_m \left[
\Theta(t)e^{- \beta \hat{H}_0}
e^{it \hat{H}} \hat{n}_{\bf k} e^{-it \hat{H}}
\right] \notag\\
&=
\sum_{\alpha,\beta,\gamma}
e^{-\beta E_\alpha ^{(m)}}
G_{\beta \alpha} ^{(m)*}(t)
\langle \psi_{\beta} ^{(m)}|\hat{n}_{\bf k}|\psi_\gamma ^{(m)}\rangle
G_{\gamma\alpha} ^{(m)}(t),
\\
Q_m 
&=
\mathrm{Tr}_m \left[ e^{- \beta \hat{H}_0} \right],
\end{align}
respectively. 
Here $m=0,1,2,\ldots$ labels the particle-number sector, $E_\alpha^{(m)}$ is the energy of the noninteracting $m$-particle state $|\psi_\alpha^{(m)}\rangle$, and $\Theta(t)$ is the unit step function. The step function enforces the post-quench evolution for $t>0$.
The propagator $G^{(m)}(t)$ describes the time evolution in the $m$-particle sector under the post-quench Hamiltonian $\hat H$, and is written as
\begin{align}
G_{\gamma\alpha} ^{(m)}(t)
=
\frac{i}{2\pi}
\int_{-\infty}^{\infty} d\omega\,
e^{-i\omega t}
G_{\gamma\alpha}^{(m)}(\omega+i0^+).
\end{align}

\begin{figure}
\centering
\includegraphics[width=0.8\textwidth]{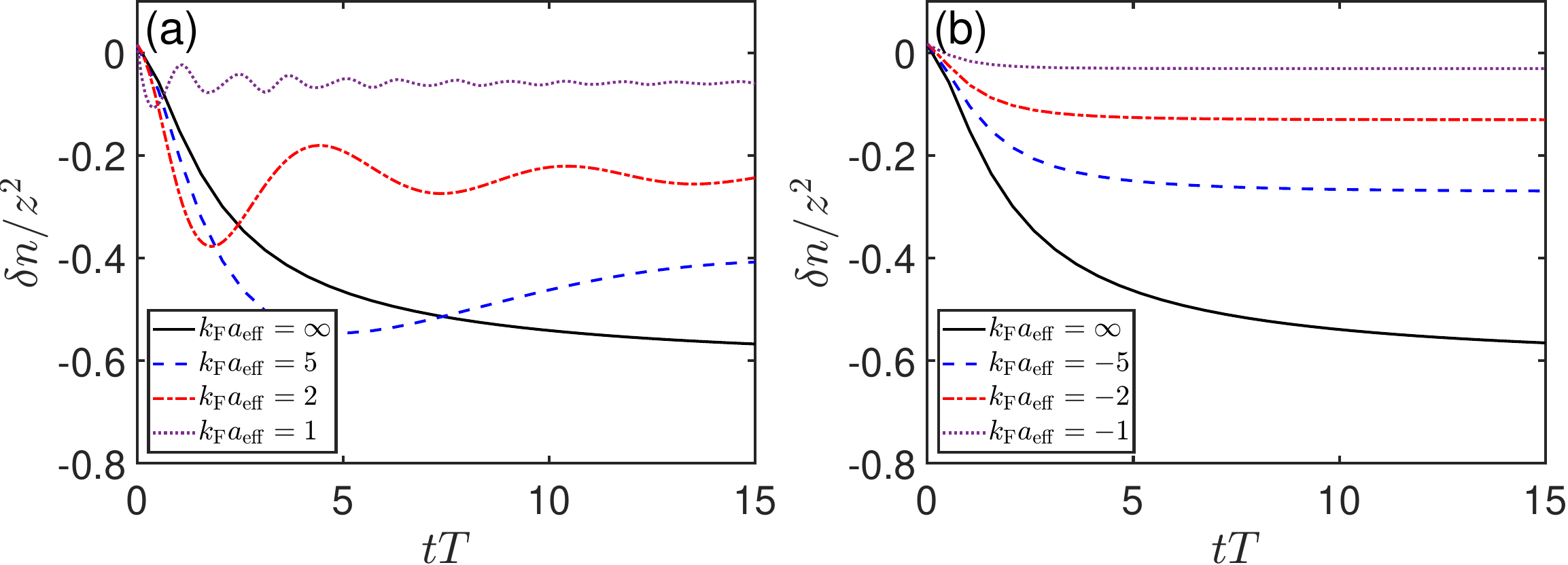}
\caption{The evolution of the momentum distribution when the quenched interaction starts from the noninteraction limit. (a): $\delta n$ evolution for four different scattering length $a_{\rm eff}>0$; (b): $\delta n$ evolution for scattering length $a_{\rm eff}<0$. Black solid line ($k_{\rm F} a_{\rm eff} =\infty$); blue dashed ($k_{\rm F} a_{\rm eff} = \pm5$); red dashed-dotted line ($k_{\rm F} a_{\rm eff} = \pm2$); purple dotted line ($k_{\rm F} a_{\rm eff} = \pm1$). The temperature of the system is $T = 0.1 T_{\rm F}$.}
\label{figS2}
\end{figure}

The high-temperature virial expansion translates the quench dynamics of the many-body system into few-body problems.
We keep the expansion up to $z^2$, which captures the leading interaction-induced two-body dynamics. For a single-particle system, $\hat H=\hat H_0$, and
\begin{align}
X_1=\mathrm{Tr}_1 \left[\Theta (t) e^{-\beta \hat H_0}e^{it\hat H_0}\hat n_{\mathbf k} e^{-it \hat H_0}\right] =e^{-\beta \frac{\mathbf k^2}{2 m_{\rm I}}}.
\end{align}
Therefore, $X_1$ and $Q_m$ are independent of time, and the time-dependent part of momentum distribution, $\delta n_{\mathbf k}=n_{\mathbf k}(t)-n_{\mathbf k}(0)$, arises only from $X_2(t)$ to order $z^2$.
Specifically,
\begin{align}
\delta n_{\mathbf k}=\left[X_2(t)-X_2(0)\right]z^2,
\label{eq:delta-n}
\end{align}
which can be obtained by solving the two-body problem.
In this case, each two-body eigenstate is labeled by two quantum numbers $| \psi _a^{(2)} \rangle =|\bf{P},\bf{q}\rangle$, with $\bf{P}$ and $\bf{q}$ being the total momentum and the relative momentum of two bosons, respectively. For the energy of the free Hamiltonian, we have
\begin{align}
E_a^{(2)} = \frac{\mathbf P^2}{4 m_{\rm I}}+\frac{\mathbf q^2}{ m_{\rm I}},
\end{align}
where $ m_{\rm I}$ is the mass of single impurity. For the two-body sector, the center-of-mass and relative motions can be separated. 
Since the interaction depends only on the relative coordinate, the nontrivial part of the two-body Green's function involves only the relative motion. 
We therefore write $\hat G^{(0)}(s)=\left(s-\varepsilon_{\mathbf q}\right)^{-1}$,
where $s=\omega+i0^+$, and $\varepsilon_{\mathbf q}=\mathbf q^2/m_{\rm I}$ is the relative kinetic energy of two impurities.
Using the Lippmann-Schwinger equation, the full two-body Green's function can be written as
\begin{align}
G_{\alpha \beta }^{(2)} &= G_{\alpha \beta }^{(0)} + G_{\alpha \beta }^{(0)}T_2(s)G_{\alpha \beta }^{(0)} \notag \\
&= \left[\frac{\langle \mathbf{q_\alpha}|\mathbf{q_\beta}\rangle}{s-\varepsilon _{\mathbf{q_\alpha}}}+\frac{T_2(s)}{(s-\varepsilon _{\mathbf{q_\alpha}})(s-\varepsilon _{\mathbf{q_\beta}})}\right]\delta_{\mathbf{P_\alpha,P_\beta}}.
\label{eq:green-single}
\end{align} 
The interaction effect entering Eq.~(\ref{eq:green-single}) is contained in the two-body scattering $T$ matrix. 
We parametrize this $T$ matrix by the effective scattering length $a_{\rm eff}$ extracted from the mediated potential $V_{\rm eff}(R,T)$,
\begin{align}
T_2(s)=\frac{4\pi / m_{\rm I}}{a_{\rm eff}^{-1}-\sqrt{-m_{\rm I}s}} .
\label{eq:T2-single}
\end{align}
Using Eqs.~(\ref{eq:green-single}) and (\ref{eq:T2-single}), we obtain the required two-body contribution $X_2$, which determines $\delta n_{\mathbf k}$ through Eq.~(\ref{eq:delta-n}).
This calculation provides a dynamical probe of the effective interaction.

In Fig.~\ref{figS2}, we show the time evolution of the lowest-momentum occupation $\delta n(k=0,t)$ after a quench from the noninteracting limit to different values of $a_{\rm eff}$.
For quenches to positive $a_{\rm eff}$, the dynamics exhibits oscillations associated with the two-body bound state.
Near the resonance, where $a_{\rm eff}^{-1}=0$, the low-momentum occupation shows a more pronounced depletion, indicating a stronger redistribution of the momentum distribution after the quench.
A similar enhancement of the low-momentum depletion is also found for quenches to negative scattering lengths near resonance.
This behavior is consistent with the resonance structure obtained from the mediated interaction.

\end{document}